\documentclass{article}
\usepackage{spconf}
\usepackage{url}
\usepackage{tipa}
\usepackage{amssymb}
\usepackage{amsmath}
\usepackage{graphicx}
\usepackage{tikz}
\usepackage[style=english]{csquotes}
\usepackage[noadjust]{cite}
\usetikzlibrary{shapes,arrows}

\title{Structure-Constrained Basis Pursuit for Compressed Sensing }
\name{Miguel Dominguez and Behnaz Ghoraani, Ph.D.}
\address{Department of Electrical Engineering \\
		 Biomedical Signal and Image Analysis Laboratory \\
		 Rochester Institute of Technology  \\
         Rochester, USA \\
         mad6384@rit.edu, bxgbme@rit.edu }
\begin{document}
\maketitle
\begin{abstract}
In compressive sensing (CS) theory, as the number of samples is decreased below a minimum threshold, the average error of the recovery increases. Sufficient sampling is either required for quality reconstruction or the error is resignedly accepted. However, most CS work has not taken advantage of the inherent structure in a variety of signals relevant to engineering applications.  Hence, this paper proposes a new method of recovery built on basis pursuit (BP), called Structure-Constrained Basis Pursuit (SCBP), that constrains signals based on known structure rather than through extra sampling. Preliminary assessments of this method on TIMIT recordings of the speech phoneme /\textipa{A}/ show a substantial decrease in error: with a fixed 5:1 compression ratio the average recovery error is 23.8\% lower versus vanilla BP. More significantly, this method can be applied to any CS application that samples structured data, such as FSK waveforms, speech, and tones.  In these cases, higher compression ratios can be reached with comparable error.
\end{abstract}
\keywords basis pursuit, compressive sensing, speech coding
\section{Introduction}
\label{intro}
In compressive sensing theory, signals that are sparse in some domain can be sampled below the Nyquist rate and still be successfully recovered using a convex optimization problem such as basis pursuit (BP) that minimizes the $\ell_1$ norm of the solution.\cite{1614066}  The number of required samples is based on the signal's sparsity and a property of the sensing matrix called \enquote{coherence} \cite{1614066,0266-5611-23-3-008,4472240}.  If the signal is sampled below this threshold, error can be introduced into the BP solution. 

Essentially, each sample is the inner product of a row of a sensing basis and the original signal.  Each sample adds a constraint to the BP problem, requiring that the final solution solve that particular inner product equation.  With enough samples, the problem is constrained enough that the correct solution is within the convex set \enquote{with overwhelming probability \cite{1614066}.}  Departing from mainstream CS discussion, this paper proposes Structure-Constrained Basis Pursuit (SCBP) as a modification to normal BP. This method tightly bounds the solution with static bounds based on structure known a priori.  Solutions to the system of equations that minimize the $\ell_1$ norm but do not look like the expected result are excluded from the final result.  Fig. \ref{fig:blockDiag} illustrates how such a system would behave compared to traditional CS.

This paper hypothesizes that such bounds can substantially reduce error when recovering an insufficiently sampled CS vector.  A speech coding application is assessed as a preliminary test.  In speech, blocks of samples are often relatively sparse in a frequency domain (such as the Discrete Cosine Transform domain) \cite{desaicompressive}. Moreover, individual phonetic elements (phonemes) can be identified by looking at a frequency-domain plot or spectrogram \cite{Rabiner:1993:FSR:153687}. Therefore, if a block of speech is guaranteed to contain only phonemes of a known spectral structure, and that structure is known a priori, SCBP exploits static upper and lower bounds defined in the frequency domain to enforce it.

\subsection{Relation to Prior Work}

The theory of compressed sensing was pioneered by Donoho, Cand\'{e}s, Romberg, Tao, and others  \cite{1614066,0266-5611-23-3-008,4472240,candes2006stable}.  The work of Boyd, Vandenberghe, Grant, and others made it possible to quickly formulate and test convex optimization problems through guidance on theory and flexible software tools \cite{Boyd:2004:CO:993483,cvx,gb08}.  Ramdas, Mishra, and Gorthi recently proposed a compressed sensing-based speech code that used the DCT as a sparsifying basis \cite{7091436}, which inspired a desire to make such a speech code even more compelling.  Yin, Morgan, Yang, and Zhang proposed a more efficient compressed sampling algorithm based on adding a rigid circulant structure to random sensing bases \cite{doi:10.1117/12.863527}. This inspired a desire to see if rigid structure in the data could also be exploited for increased performance.\linebreak

Section \ref{theory} describes the mathematics of compressive sensing and recovery.  Section \ref{sec:constrainedBP} introduces the SCBP method.  Section \ref{sec:exper} provides the experimental setup testing the technique on a single phoneme, /\textipa{A}/.  Section \ref{sec:experResults} discusses the results of the experiment. Section \ref{sec:conclusion} is the conclusion and Section \ref{sec:Ack} shows acknowledgements. 

\section{Compressive Sensing Theory}
\label{theory}
\tikzstyle{block} = [draw, rectangle, 
    minimum height=3em, minimum width=2em]
\tikzstyle{sum} = [draw, circle, node distance=1cm]
\tikzstyle{input} = [coordinate]
\tikzstyle{output} = [coordinate]
\tikzstyle{pinstyle} = [pin edge={to-,thin,black}]

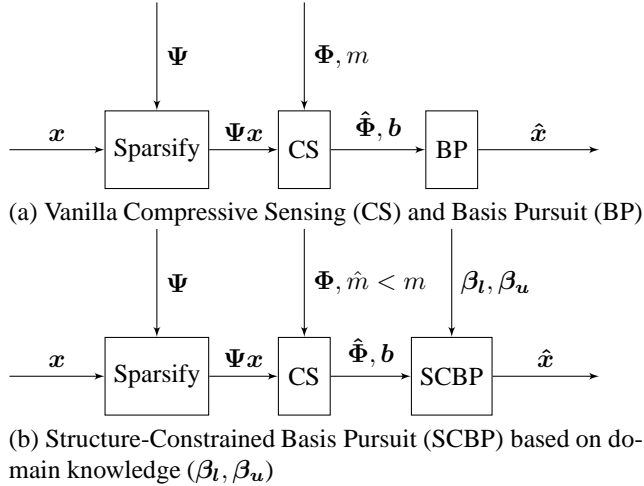
\begin {figure}
\centering
  \resizebox {\linewidth} {!} {
  \begin{minipage}[b]{1.0\linewidth}
   \begin{tikzpicture}[auto, node distance=2cm,>=latex']
        \node [input, name=signal] {};
        \node [block, right of=signal] (sparsify) {Sparsify};
        \node [input, above of=sparsify,name=psi] {};
        \node [block, right of=sparsify] (cs) {CS};
        \node [input, above of=cs, name=phi] {};
        \node [block, right of=cs] (bp) {BP};
        \node [output, right of=bp, name=xhat] {};
        \draw [->] (sparsify) -- node[name=psiX] {$\boldsymbol{\Psi x}$} (cs);
        \draw [->] (cs) -- node[name=phiPsiX] {$\boldsymbol{\hat{\Phi}},\boldsymbol{b}$} (bp);
        \draw [draw,->] (signal) -- node[] {$\boldsymbol{x}$} (sparsify);
        \draw [draw,->] (psi) -- node[] {$\boldsymbol{\Psi}$} (sparsify);
        \draw [draw,->] (phi) -- node[] {$\boldsymbol{\Phi}, m$} (cs);
        \draw [draw,->] (bp) -- node[] {$\boldsymbol{\hat{x}}$} (xhat);
    \end{tikzpicture}
    
    (a) Vanilla Compressive Sensing (CS) and Basis Pursuit (BP)
   \end{minipage}
     }
   \resizebox {\linewidth} {!} {
   \begin{minipage}[b]{1.0\linewidth}
   \begin{tikzpicture}[auto, node distance=2cm,>=latex']
        \node [input, name=signal] {};
        \node [block, right of=signal] (sparsify) {Sparsify};
        \node [input, above of=sparsify,name=psi] {};
        \node [block, right of=sparsify] (cs) {CS};
        \node [input, above of=cs, name=phi] {};
        \node [block, right of=cs] (bp) {SCBP};
        \node [output, right of=bp, name=xhat] {};
        \node [input, above of=bp, name=domain] {};
        \draw [->] (sparsify) -- node[name=psiX] {$\boldsymbol{\Psi x}$} (cs);
        \draw [->] (cs) -- node[name=phiPsiX] {$\boldsymbol{\hat{\Phi}},\boldsymbol{b}$} (bp);
        \draw [draw,->] (signal) -- node[] {$\boldsymbol{x}$} (sparsify);
        \draw [draw,->] (psi) -- node[] {$\boldsymbol{\Psi}$} (sparsify);
        \draw [draw,->] (phi) -- node[] {$\boldsymbol{\Phi}, \hat{m} < m$} (cs);
        \draw [draw,->] (bp) -- node[] {$\boldsymbol{\hat{x}}$} (xhat);
        \draw [draw,->] (domain) -- node[] {$\boldsymbol{\beta_l},\boldsymbol{\beta_u}$} (bp);
    \end{tikzpicture}
    
    (b) Structure-Constrained Basis Pursuit (SCBP) based on domain knowledge ($\boldsymbol{\beta_l},\boldsymbol{\beta_u}$)
   \end{minipage}
   }
\caption{Compressive sensing and recovery with basis pursuit. Domain knowledge can reduce the required $m$ and still maintain reasonable error.}
\label{fig:blockDiag}
\end{figure}
Compressive sensing concerns the sampling of sparse signals.  A signal $\boldsymbol{x} \in \mathbb{R}^{n} $ is called $S$-sparse if all but $S$ of its elements are zero or near-zero \cite{4472240}.  Many signals are not naturally sparse but can be represented sparsely by a linear transformation with an orthonormal \enquote{sparsifying} basis $\boldsymbol{\Psi}$.  The Discrete Cosine Transform (DCT) basis is one such sparsifying basis, used for this purpose in practical compression schemes such as JPEG \cite{125072}.

Conventional sampling involves direct, evenly-spaced measurements of a signal.  It can be modeled, roughly, as in Eq. (\ref{naiveSample}) \cite{4472240}.
\begin{equation}
\label{naiveSample}
\boldsymbol{Ix} = \boldsymbol{b}
\end{equation}
The identity matrix $\boldsymbol{I}$ can be thought of as a \enquote{sensing} basis composed of time-shifted impulses.  The resulting sampled signal is $\boldsymbol{b}$.

Compressed sensing theory posits that a sensing basis $\boldsymbol{\Phi}\in\mathbb{R}^{n\times n}$ could be defined such that not every row of the basis needs to be used in the sampling equation.  A matrix $\boldsymbol{\hat{\Phi}}\in\mathbb{R}^{m\times n}$ can be formed by a subset of $m<<n$ rows of this basis, and sampling is performed as in Eq. (\ref{compressedSample}).
\begin{equation}
\label{compressedSample}
\boldsymbol{\hat{\Phi}x} = \boldsymbol{b}, \boldsymbol{b}\in\mathbb{R}^{m}
\end{equation}
The sensing basis should be chosen so that correlation is minimized between any given row of the sensing basis $\boldsymbol{\Phi}$ and any given row of the sparsifying basis $\boldsymbol{\Psi}$.  The worst-case measure of this row-wise correlation is typically referred to as \enquote{coherence} \cite{4472240}.  Typically, sensing bases of Gaussian or Bernoulli random numbers decently minimize this correlation \cite{1614066,candes2006stable}.

The resulting vector $\boldsymbol{b}$ is not usable in its present form.  It needs to be \enquote{recovered} into an approximation of the original signal.  To recover the original, Eq. (\ref{compressedSample}) is posed again, with $\boldsymbol{x}$ as the unknown quantity and $\boldsymbol{b}$ as the known quantity.  This is an underdetermined linear system of equations, likely having multiple solutions. A measure based on the solution vector should be minimized or maximized to choose the optimal solution.  The signal $\boldsymbol{\Psi x} $ is known to be sparse, so this problem could be posed as a convex optimization problem designed to maximize sparsity.  Unfortunately, optimizing for the $\ell_0$ \enquote{norm,} a count of the number of nonzero elements in the vector, is a combinatorial problem \cite{Elad:2010:SRR:1895005}.  However, minimizing the $\ell_1$ norm of the solution (also known as \enquote{basis pursuit} or BP) also promotes sparsity, and is convex \cite{1614066,Boyd:2004:CO:993483}.  Eq. (\ref{compressedBP}) poses BP in terms of a CS application.
\begin{equation}
\label{compressedBP}
\text{minimize} \;||\boldsymbol{\Psi x}||_1\;
\text{subject to}
\; \boldsymbol{\hat{\Phi}}\boldsymbol{\Psi x} = \boldsymbol{b}\;
\end{equation}
The minimum $m$ required to effectively recover $\boldsymbol{x}$ is a function of the sparsity and the coherence \cite{0266-5611-23-3-008}.  The sparser $\boldsymbol{\Psi x}$ is and the more incoherent $\boldsymbol{\Psi}$ and $\boldsymbol{\Phi}$ are, the fewer samples are required.  If not enough samples are taken, error between the desired signal and the recovered signal will increase.  Essentially, too few samples mean too few constraints in the basis pursuit problem, and minimizing the $\ell_1$ norm will not necessarily lead to the desired solution.
\section{The Proposed Structure-Constrained Basis Pursuit}
\label{sec:constrainedBP}
This paper puts forward a modification to the basis pursuit formulation shown in Eq. (\ref{compressedBP}).  If an insufficient number of samples are taken, but the sparse solution vector can be bounded between an upper bound $\alpha \boldsymbol{\beta_u}$ and a lower bound $\alpha \boldsymbol{\beta_l}$ ($\boldsymbol{\beta_u},\boldsymbol{\beta_l},\boldsymbol{x}\in\mathbb{R}^{n}$ and $\alpha\in\mathbb{R}$ is a scale factor), then a superior recovery with reduced error shall be found in Eq. (\ref{compressedCBP}).
\begin{equation}
\label{compressedCBP}
\begin{split}
\text{minimize} \;||\boldsymbol{\Psi x}||_1 + ||\alpha ||_1\; \\
\text{subject to}
\; ||\boldsymbol{\hat{\Phi}}\boldsymbol{\Psi x} - \boldsymbol{b}||_1 < \epsilon \; \\
\text{and}\; \alpha \boldsymbol{\beta_l} \preceq \boldsymbol{\Psi} x \preceq \alpha \boldsymbol{\beta_u}
\end{split}
\end{equation}

This Structure-constrained Basis Pursuit explicitly forbids solutions that minimize the $\ell_1$ norm, but do not have a comparable structure to the expected signal. The structure is enforced using upper and lower bounds $\boldsymbol{\beta_l}$ and $\boldsymbol{\beta_u}$.  The use of upper and lower bounds has been suggested previously for regression problems by Boyd and Vandenberghe \cite{Boyd:2004:CO:993483}. The requirement of the compressed sensing equation being exactly solved can be loosened by restricting the solution's inexactitude to within a handpicked $\epsilon$.  The scale factor $\alpha$ is optimized so that the magnitude of the bounds are sized to encompass the signal as closely as possible.

For speech, whose phonemes offer predictable frequency characteristics regardless of the speaker, tight bounds $\boldsymbol{\beta_l}$ and $\boldsymbol{\beta_u}$ should be able to be constructed in terms of the signal's DCT-domain representation.  If the phoneme could somehow be known or surmised, tight bounds of this form would enable SCBP to reduce recovery error in the absence of sufficient compressive sampling.  The loudness or softness of the speech is irrelevant, as the optimal scale of the bounds will be discovered by the convex solver.

\section{Experimental Setup}
\label{sec:exper}
To evaluate SCBP, this paper samples and recovers a subset of TIMIT \cite{TIMIT}.  All instances of /\textipa{A}/ alone are extracted into individual vectors and separated into test and training sets according to TIMIT's \enquote{TEST} and \enquote{TRAIN} partitions, respectively.  This is trivial because TIMIT provides phonetic transcripts that define the beginning and ending sample for each phoneme in a recording.  Vectors longer than 1024 samples are cut into 1024-sample blocks and a remainder block, in order to limit the size of the basis pursuit problem.  This means that an 1025-sample vector is split into a 1024-sample block and a 1-sample block.  In this dataset, 59\% of the vectors are exactly 1024 samples long and only 3.8\% have a length less than or equal to 100 samples.

All of the vectors are sampled at 16 kHz.  Many practical real-time speech codes impose an 8 kHz sample rate, because much of the speech frequency content used is concentrated between 300-3400 Hz \cite{Huang:2001:SLP:560905}.  However, some of the higher-frequency content (such as fricative data) is lost at this sampling rate.  This test is keeping the higher sampling rate in order to maintain fidelity of the signal and because speech data should become relatively more sparse as sampling rate increases.  Perception of speech content is minimally affected by frequencies beyond 10 kHz \cite{Huang:2001:SLP:560905}.

The phoneme /\textipa{A}/ was chosen for study because it is a relatively orderly phoneme:  It is a vowel, which means that instances of it tend to be long, periodic, and sparse in the DCT domain. That said, experimental evidence shows that the formant frequencies that identify a vowel can vary widely per person \cite{Rabiner:1993:FSR:153687}.  That effective bounds can be decided in the face of this fact shall be demonstrated by this paper.

\begin{figure}[htb]
\includegraphics[width=8.5cm]{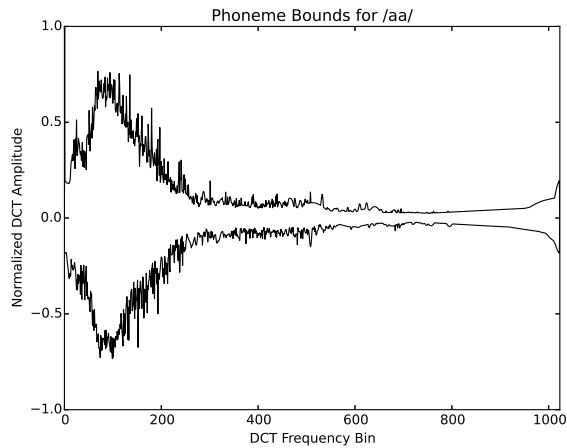}
\caption{Phoneme bounds for /\textipa{A}/.}
\label{fig:bounds}
\end{figure}

Upper and lower bounds were constructed by taking each /\textipa{A}/ vector from TIMIT's \enquote{TRAIN} set, transforming it with SciPy's orthonormal DCT function, interpolating it to exactly 1024 samples with SciPy's \enquote{resample} function, and normalizing using the $\ell_2$ norm. This produces row vectors $\gamma_i^\intercal$ ($i=1,2,\dots N$ for $N$ training vectors).  The processed vectors are stacked into a matrix $\boldsymbol{\Gamma}\in\mathbb{R}^{N\times 1024}$ and the maximum and minimum of each column are the upper and lower bound for the corresponding element in a vector $\boldsymbol{x}\in\mathbb{R}^{1024}$, as shown in Eq. (\ref{eqbeta}).
\begin{equation}
\begin{split}
\boldsymbol{\beta_u}(j) = \max_i \boldsymbol{\Gamma}[i,j] \\
\boldsymbol{\beta_l}(j) = \min_i \boldsymbol{\Gamma}[i,j] \\
j = 1,2,\ldots ,1024
\end{split}
\label{eqbeta}
\end{equation}
These are the most inclusive bounds based on the training data.  Fig. \ref{fig:bounds} shows the bounds defined for /\textipa{A}/.  There is a clear structure, with low frequencies containing most of the energy and high frequencies much less represented.  This follows general speech patterns, where most energy is concentrated in low frequencies.  For test vectors that are less than 1024 samples, these bounds are decimated with MATLAB's resample command to the correct length.  These decimated bounds will still be useful, as lower-resolution DCTs maintain the same general shape.  The only difference is that each \enquote{bin} in a DCT with fewer samples will represent a wider range of frequencies.

The entire set of \enquote{TEST} /\textipa{A}/ vectors were compressively sampled at a fixed CR of 5:1, 100 times each for vanilla BP and just over 100 times each for SCBP.  Data collection errors in one of the constrained runs led to it being thrown out and rerun, resulting in exactly 100 runs consisting of 278,000 total data points for each type of BP.  In both cases, $\boldsymbol{\Psi}$ was the DCT basis and $\boldsymbol{\Phi}$ was Gaussian random with zero mean and unit variance, orthonormalized with QR Factorization \cite{Boyd:2004:CO:993483}. In the SCBP tests, $\epsilon$ was set to 0.001.  The compressive samples were recovered using CVX with MATLAB \cite{cvx,gb08}.

\section{Experimental Results}
\label{sec:experResults}
Recovery error is measured in normalized mean squared error (NMSE), which is MSE divided by the mean energy of the original signal, as shown in Eq. (\ref{nmse}).

\begin{equation}
NMSE = \frac{\sum\limits_{k=1}^n(x[k] - \hat{x}[k])^2}{\sum\limits_{k=1}^n x[k]^2}
\label{nmse}
\end{equation}

\begin{figure}[htb]
\includegraphics[width=8.5cm]{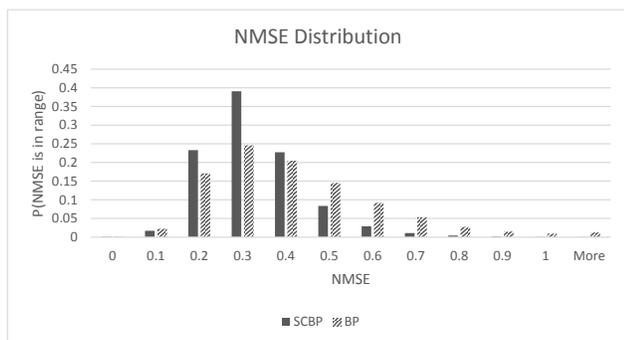}
\caption{Distribution of error with BP and with SCBP.}
\label{histogramResults}
\end{figure}

The data from the evaluation clearly favors SCBP in this application. Fig. \ref{histogramResults} compares SCBP to BP.  SCBP's distribution of error has a 23.8\% lower average than BP, a 65.7\% smaller variance, and a skew weighted toward lower error.  Table \ref{statsResults} displays BP and SCBP NMSE means and variances. The data shows that SCBP prevents many solutions that misrepresent the sampled signal despite minimizing the $\ell_1$ norm using standard BP.

CVX was unable to recover one of the 278,000 SCBP problems.  This data point is not included in the mean/variance calculations, and in Fig. \ref{histogramResults} this data point is sorted under the "More" error bin.

\begin{table}[htb]
\centering
\resizebox{8.5cm}{!}{%
\begin{tabular}{l|l|l|l|}
\cline{2-4}
                                    & BP      & SCBP        & \% Improvement \\ \hline
\multicolumn{1}{|l|}{NMSE Mean}     & 0.367   & 0.280       & 23.8\%    \\ \hline
\multicolumn{1}{|l|}{NMSE Variance} & 0.040   & 0.014       & 65.7\%   \\ \hline
\end{tabular}
}
\caption{Statistical performance over all /\textipa{A}/ data.}
\label{statsResults}
\end{table}

With the extra constraints and the requirement that $\alpha$ be optimized as part of the BP operation, SCBP is considerably slower than BP.  Actual timing numbers are not presented in this paper due to the fact that the speed of the machines used to perform the test was not controlled, but there was roughly a six-fold increase in recovery time with SCBP vs vanilla BP. Future work could determine if $\alpha$ could be efficiently computed beforehand based on knowledge of the BP inputs in order to free the convex solver from the task.

\section{Conclusion}
\label{sec:conclusion}

This paper proposed Structure-Constrained Basis Pursuit (SCBP) as a method of recovering a signal from insufficient compressed samples with lower error. While typical CS applications build BP constraints solely from compressive samples, SCBP enables static constraints based on the signal's structure to act as an effective substitute.  A small speech-related testbed demonstrated compelling improvements of 23.8\% in recovery error due to SCBP.  This provides numerous opportunities for future work.  Discovering the structure of a signal before BP is an open question. Compressed samples measured with an appropriate sensing basis could betray some structure of the original signal.  Hidden Markov Models and other speech modeling techniques could be applied to predict the structure of compressed speech.  In a future work, the authors intend to pursue the exploitation of CS basis structure for feature extraction to enable SCBP in practical speech coding.

However, SCBP has a much wider application than as a method for speech coding.  Any compressed sensing application where the data is predictably structured can benefit from this technique, resulting in lower sampling rates that still acceptably recover data.  Two obvious examples are tone detectors and FSK receivers, where the problem can be tightly constrained to only seek a narrow range of frequencies.  Transforms besides frequency domain transforms may also be used to great effect.  While bound constraints have been a feature of convex solvers for some time, there exists substantial opportunity to exploit them through SCBP for improved sampling.

\section{Acknowledgements}
\label{sec:Ack}
The authors would like to thank Dr. Ernest Fokoue for his assistance with convex optimization concepts and his insight that helped increase the performance of SCBP.
\bibliography{csspeech}
\bibliographystyle{IEEE}
\end{document}